\documentclass[ejp,preprint]{revtex4-1}

\usepackage{amssymb}
\usepackage{amsmath}

\begin{document}

\title{A Laplace transform approach\\ to the quantum harmonic oscillator\footnote{To appear in European Journal of Physics}}
\author{Douglas R. M. Pimentel and Antonio S. de Castro}
\email{Author to whom correspondence should be addressed. Electronic mail:
castro@pq.cnpq.br}
\affiliation{Universidade Estadual Paulista, Faculdade de Engenharia de
Guaratinguet\'{a}, Departamento de F\'{i}sica e Qu\'{i}mica, 12516-410
Guaratinguet\'{a} SP - Brazil}

\begin{abstract}
The one-dimensional quantum harmonic oscillator problem is examined via the
Laplace transform method. The stationary states are determined by requiring
definite parity and good behaviour of the eigenfunction at the origin and at
infinity.
\end{abstract}

\maketitle

\section{Introduction}

The quantum harmonic oscillator is one of the most important systems in
quantum mechanics because it can be solved in closed form and its solution
can be useful as approximations or exact solutions of various problems. The
harmonic oscillator is usually solved with the help of the power series
method \cite{som}, by the algebraic method based on the algebra of operators
\cite{sak}, or by the path integral approach \cite{fey}. In recent times,
the one-dimensional harmonic oscillator has also been approached by the
Fourier transform method \cite{mun}-\cite{pal}. \ Another operational method
useful in quantum mechanics is the Laplace transform method. The Laplace
transform method was used at the first years of quantum mechanics by Schr%
\"{o}dinger into the discussion of radial eigenfunction of the hydrogen atom
\cite{sch}, and more than forty years later Englefield approached the Schr%
\"{o}dinger equation with the Coulomb, oscillator, exponential and Yamaguchi
potentials \cite{eng}. More than twenty years went by and the hydrogen atom
was again examined with the Laplace transform method \cite{swa}. For some
years now the $1/x$ \cite{ran}, Morse \cite{che1}, $N$-dimensional harmonic
oscillator \cite{che2}, pseudoharmonic and Mie-type \cite{ard}, and Dirac
delta \cite{asc} potentials have been solved for the Laplace transform.

In Ref.  \cite{eng}, Englefield found the spectrum of the three-dimensional
harmonic oscillator by imposing that the radial eigenfunction vanishes at
the origin and by using the closed-form solution for the Laplace transform.
This paper approaches the one-dimensional Schr\"{o}dinger equation for the
harmonic oscillator with the Laplace transform method following the recipe
proposed by Englefield \cite{eng}. Nevertheless, we do not use the
closed-form solution for the Laplace transform. In addition, we enlarge the
class of problems to include eigenfunctions satisfying homogeneous Neumann
condition at the origin. The main features of our approach are as follows.
After factorizing the behaviour at infinity, the second-order differential
equation for the eigenfunction $\psi \left( x\right) $ transmutes in a
nonhomogeneous first-order equation for the Laplace transform $\Phi \left(
s\right) $. The closed-form solution for $\Phi \left( s\right) $ is not
necessary. After inspecting the singularities of the differential equation
for $\Phi \left( s\right) $, we make a series expansion about the
appropriate singular point, related to the proper behaviour of $\psi \left(
x\right) $ near infinity. The behaviour of the eigenfunction near the
origin, related to the behaviour of $\Phi \left( s\right) $ near infinity,
implies to truncate the series. The root of the indicial equation is the
quantization condition. The recurrence relation for the coefficients of the
truncated series for $\Phi \left( s\right) $ is solved and the inverse
Laplace transform is identified with the Hermite polynomials. This procedure
should be helpful for students in a course on mathematical physics.

\section{The Laplace transform applied to the harmonic oscillator}

Let us begin with a brief description of the Laplace transform and a few of
its properties \cite{doe}. The Laplace transform of a function $f(t)$ is
defined by
\begin{equation}
F(s)=\mathcal{L}\left\{ f\left( t\right) \right\} =\int_{0}^{\infty
}dt\,e^{-st}f\left( t\right) .  \label{l1}
\end{equation}%
If there is some constant $\sigma \in
\mathbb{R}
$ such that
\begin{equation}
|e^{-\sigma t}f\left( t\right) |\leq M,  \label{l2}
\end{equation}%
for sufficiently large $t$, the integral in Equation (\ref{l1}) will exist
for \textrm{Re}$\,s>\sigma $ and $f\left( t\right) $ is said to be of
exponential order. The Laplace transform may fail to exist because of a
sufficiently strong singularity in the function $f\left( t\right) $ as $%
t\rightarrow 0$. In particular%
\begin{equation}
\mathcal{L}\left\{ \frac{t^{\alpha }}{\Gamma \left( \alpha +1\right) }%
\right\} =\frac{1}{s^{\alpha +1}},\quad \alpha >-1.  \label{l7}
\end{equation}%
The Laplace transform has the derivative properties%
\begin{eqnarray}
\mathcal{L}\left\{ f^{\left( n\right) }\left( t\right) \right\} &=&s^{n}%
\mathcal{L}\left\{ f\left( t\right) \right\}
-\sum\limits_{k=0}^{n-1}s^{n-1-k}f^{\left( k\right) }\left( 0\right)  \notag
\\
&&  \label{l3} \\
\mathcal{L}\left\{ t^{n}f\left( t\right) \right\} &=&\left( -1\right)
^{n}F^{\left( n\right) }\left( s\right) ,  \notag
\end{eqnarray}%
where the superscript $(n)$ stands for the \textit{n}-th derivative with
respect to $t$ for $f^{\left( n\right) }\left( t\right) $, and with respect
to $s$ for $F^{\left( n\right) }\left( s\right) $. If near a singular point $%
s_{0}$ the Laplace transform behaves as%
\begin{equation}
F\left( s\right) \underset{s\rightarrow s_{0}}{\sim }\frac{1}{\left(
s-s_{0}\right) ^{\nu }}  \label{l4}
\end{equation}%
then
\begin{equation}
f\left( t\right) \underset{t\rightarrow \infty }{\sim }\frac{1}{\Gamma
\left( \nu \right) }\,t^{\nu -1}\,e^{s_{0}t},  \label{l5}
\end{equation}%
where $\Gamma \left( \nu \right) $ is the gamma function. On the other hand,
if near the origin $f\left( t\right) $ behaves like $t^{\alpha }$, with $%
\alpha >-1$, then $F\left( s\right) $ behaves near infinity as%
\begin{equation}
F\left( s\right) \underset{s\rightarrow \infty }{\sim }\,\frac{\Gamma \left(
\alpha +1\right) }{s^{\alpha +1}}.  \label{l6}
\end{equation}

We are now prepared to address the quantum harmonic oscillator problem. The
one-dimensional Schr\"{o}dinger equation for the harmonic oscillator reads
\begin{equation}
\left( -\frac{\hslash ^{2}}{2m}\frac{d^{\,2}}{dx^{2}}+\frac{1}{2}m\omega
^{2}x^{2}-E\right) \psi \left( x\right) =0  \label{eq1}
\end{equation}%
Because the harmonic oscillator potential is invariant under reflection
through the origin ($x\rightarrow -x$), eigenfunctions with well-defined
parities can be built. Thus, it suffices to concentrate attention on the
positive half-line and impose boundary conditions on $\psi $ at the origin
and at infinity. Normalizability demands $\psi \rightarrow 0$ as $%
x\rightarrow \infty $. Eigenfunctions and their first derivatives continuous
on the whole line with well-defined parities can be constructed by taking
symmetric and antisymmetric linear combinations of $\psi $ defined on the
positive side of the $x$-axis. As $x\rightarrow 0$, the solution varies as $%
x^{\delta }$, where $\delta $ is $0$ or $1$. The homogeneous Neumann
condition ($\left. d\psi /dx\right\vert _{x=0}=0$), develops for $\delta =0$
but not for $\delta =1$ whereas the homogeneous Dirichlet boundary condition
($\psi \left( 0\right) =0$) develops for $\delta =1$ but not for $\delta =0$%
. The continuity of $\psi $ at the origin excludes the possibility of an
odd-parity eigenfunction for $\delta =0$, and the continuity of $d\psi /dx$
at the origin excludes the possibility of an even-parity eigenfunction for $%
\delta =1$. Thus,%
\begin{equation}
\psi \left( x\right) \underset{x\rightarrow 0}{\sim }x^{\delta },\quad
\delta =\left\{
\begin{array}{c}
0 \\
\\
1%
\end{array}%
\begin{array}{c}
\mathrm{for}\text{ }\psi \text{ \textrm{even}} \\
\\
\mathrm{for}\text{ }\psi \text{ \textrm{odd.}}%
\end{array}%
\right.  \label{eqa}
\end{equation}%
On the other hand, the normalizable asymptotic form of the solution as $%
x\rightarrow \infty $ is given by%
\begin{equation}
\psi \left( x\right) \underset{x\rightarrow \infty }{\sim }e^{-m\omega
x^{2}/(2\hslash )}.
\end{equation}%
This behaviour invites us to define $\xi =m\omega x^{2}/\hslash $ in such a
way that Equation (\ref{eq1}) is written as
\begin{equation}
\,\left( \xi \frac{d^{\,2}}{d\xi ^{2}}+\frac{1}{2}\frac{d}{d\xi }+\frac{%
k-\xi }{4}\right) \psi \left( \xi \right) =0,  \label{eq3}
\end{equation}%
\noindent where $k=2E/\left( \hslash \omega \right) $. The solution for all $%
\xi $ can now be written as%
\begin{equation}
\psi \left( \xi \right) =\phi \left( \xi \right) e^{-\xi /2},  \label{eq4}
\end{equation}%
where the unknown $\phi \left( \xi \right) $ is solution of the confluent
hypergeometric equation \cite{leb}
\begin{equation}
\xi \,\frac{d^{\,2}\phi \left( \xi \right) }{d\xi ^{2}}+(b-\xi )\,\frac{%
d\phi \left( \xi \right) }{d\xi }-a\,\phi \left( \xi \right) =0,  \label{kum}
\end{equation}%
\noindent with $a=\left( 1-k\right) /4$ and $b=1/2$. Because the asymptotic
behaviour of \ $\psi \left( \xi \right) $ is given by $\exp (-\xi /2)$ as $%
\xi \rightarrow \infty $, one has to find a particular solution of (\ref{kum}%
) in such a way that $\phi \left( \xi \right) $ tends to infinity no more
rapidly than $\exp \left( \alpha \xi ^{\beta }\right) $, with $\beta <1$ and
arbitrary $\alpha $, for sufficiently large $\xi $. This occurs because $%
\alpha \xi ^{\beta }-\xi /2\rightarrow -\xi /2$ as $\xi \rightarrow \infty $%
. This condition, added by the fact that $\phi \left( \xi \right) $ varies
near the origin as $\xi ^{\delta /2}$, ensures the existence of the Laplace
transform of $\phi \left( \xi \right) $.

Denoting $\Phi (s)=\mathcal{L}\left\{ \phi \left( \xi \right) \right\} $ and
using the derivative properties of the Laplace transform given by Eq. (\ref%
{l3}), the transform of Eq. (\ref{kum}) furnishes the nonhomogeneous
first-order differential equation for $\Phi \left( s\right) $:
\begin{equation}
s\left( s-1\right) \frac{d\Phi \left( s\right) }{ds}+\left( \frac{3}{2}\,s-%
\frac{k+3}{4}\right) \Phi \left( s\right) =\frac{\phi \left( 0\right) }{2}.
\label{EQphi}
\end{equation}%
Notice that $s=0$ and $s=1$ are singular points of this differential
equation. To make use of the property of the Laplace transform near a
singular point, Eqs. (\ref{l4}) and (\ref{l5}), and taking into account the
asymptotic behaviour of $\phi \left( \xi \right) $ near infinity, we try a
series expansion of $\Phi \left( s\right) $ about $s=0$ \cite{f1}:
\begin{equation}
\Phi _{\nu }\left( s\right) =s^{-\nu }\sum_{j=0}^{\infty }c_{j}^{\left( \nu
\right) }\,s^{j},\quad c_{0}^{\left( \nu \right) }\neq 0.  \label{phi}
\end{equation}%
Referring to Eqs. (\ref{l6}) and (\ref{eqa}), we find
\begin{equation}
\Phi \left( s\right) \underset{s\rightarrow \infty }{\sim }\left\{
\begin{array}{c}
\frac{\Gamma \left( 1\right) }{s} \\
\\
\frac{\Gamma \left( 3/2\right) }{s^{3/2}}%
\end{array}%
\begin{array}{c}
\mathrm{for}\text{ }\psi \text{ \textrm{even}} \\
\\
\mathrm{for}\text{ }\psi \text{ \textrm{odd.}}%
\end{array}%
\right.
\end{equation}%
Thus, the series terminates at $j=n$ in such a way that%
\begin{equation}
\nu =\left\{
\begin{array}{c}
n+1 \\
\\
n+3/2%
\end{array}%
\begin{array}{c}
\mathrm{for}\text{ }\psi \text{ \textrm{even}} \\
\\
\mathrm{for}\text{ }\psi \text{ \textrm{odd}}%
\end{array}%
\right.
\end{equation}%
and
\begin{equation}
c_{n}^{\left( n\right) }=\left\{
\begin{array}{c}
\phi \left( 0\right) \\
\\
\mathrm{arbitrary}%
\end{array}%
\begin{array}{c}
\mathrm{for}\text{ }\psi \text{ \textrm{even}} \\
\\
\mathrm{for}\text{ }\psi \text{ \textrm{odd.}}%
\end{array}%
\right.
\end{equation}%
Beyond to make $\phi \left( \xi \right) $ to behave as $\xi ^{\nu -1}$ as $%
\xi \rightarrow \infty $, another important consequence of the term $s^{-\nu
}\,c_{0}^{\left( \nu \right) }$ in Eq. (\ref{phi}) is to give rise to the
quantization condition $\nu =(k+3)/4$ \cite{f2}. Thus,
\begin{equation}
E_{n}=\hslash \omega \times \left\{
\begin{array}{c}
2n+1/2 \\
\\
2n+1+1/2%
\end{array}%
\begin{array}{c}
\mathrm{for}\text{ }\psi \text{ \textrm{even}} \\
\\
\mathrm{for}\text{ }\psi \text{ \textrm{odd.}}%
\end{array}%
\right.
\end{equation}%
Inserting Eq. (\ref{phi}) into Eq. (\ref{EQphi}) we obtain the recurrence
relation%
\begin{equation}
c_{j+1}^{\left( n\right) }=-\frac{c_{j}^{\left( n\right) }}{j+1}\times
\left\{
\begin{array}{c}
n-j+1/2 \\
\\
n-j%
\end{array}%
\begin{array}{c}
\mathrm{for}\text{ }\psi \text{ \textrm{even}} \\
\\
\mathrm{for}\text{ }\psi \text{ \textrm{odd.}}%
\end{array}%
\right.
\end{equation}%
Inspection and induction yields
\begin{equation}
c_{j}^{\left( n\right) }=c_{0}^{\left( n\right) }\frac{\left( -1\right) ^{j}%
}{j!}\times \left\{
\begin{array}{c}
\frac{\Gamma \left( n+1/2\right) }{\Gamma \left( n-j+1/2\right) } \\
\\
\frac{n!}{\left( n-j\right) !}%
\end{array}%
\begin{array}{c}
\mathrm{for}\text{ }\psi \text{ \textrm{even}} \\
\\
\mathrm{for}\text{ }\psi \text{ \textrm{odd,}}%
\end{array}%
\right.
\end{equation}%
so that%
\begin{equation}
\Phi _{n}\left( s\right) =c_{0}^{\left( n\right) }\times \left\{
\begin{array}{c}
\Gamma \left( n+1/2\right) \sum_{j=0}^{n}\frac{\left( -1\right) ^{j}}{%
j!\Gamma \left( n-j+1/2\right) s^{n-j+1}} \\
\\
n!\sum_{j=0}^{n}\frac{\left( -1\right) ^{j}}{j!\left( n-j\right) !s^{n-j+3/2}%
}%
\end{array}%
\begin{array}{c}
\mathrm{for}\text{ }\psi \text{ \textrm{even}} \\
\\
\mathrm{for}\text{ }\psi \text{ \textrm{odd.}}%
\end{array}%
\right.
\end{equation}%
Using Eq. (\ref{l7}) and inverting the Laplace transform term by term, we
can reconstruct $\phi _{n}\left( \xi \right) $:%
\begin{equation}
\phi _{n}\left( \xi \right) =c_{0}^{\left( n\right) }\times \left\{
\begin{array}{c}
\Gamma \left( n+1/2\right) \sum_{j=0}^{n}\frac{\left( -1\right) ^{j}\xi
^{n-j}}{j!\Gamma \left( n-j+1\right) \Gamma \left( n-j+1/2\right) } \\
\\
n!\sum_{j=0}^{n}\frac{\left( -1\right) ^{j}\xi ^{n-j+1/2}}{j!\left(
n-j\right) !\Gamma \left( n-j+3/2\right) }%
\end{array}%
\begin{array}{c}
\mathrm{for}\text{ }\psi \text{ \textrm{even}} \\
\\
\mathrm{for}\text{ }\psi \text{ \textrm{odd.}}%
\end{array}%
\right.
\end{equation}%
Using Legendre's duplication formula \cite{leb}
\begin{equation}
\Gamma \left( z+1\right) \Gamma \left( z+1/2\right) =2^{-2z}\sqrt{\pi }%
\,\Gamma \left( 2z+1\right) ,  \label{le}
\end{equation}%
$\phi _{n}\left( \xi \right) $ turns out to be%
\begin{equation}
\phi _{n}\left( \xi \right) =\frac{c_{0}^{\left( n\right) }}{\sqrt{\pi }}%
\times \left\{
\begin{array}{c}
\Gamma \left( n+1/2\right) \sum_{j=0}^{n}\frac{\left( -1\right) ^{j}\left( 2%
\sqrt{\xi }\right) ^{2n-2j}}{j!\left( 2n-2j\right) !} \\
\\
n!\sum_{j=0}^{n}\frac{\left( -1\right) ^{j}\left( 2\sqrt{\xi }\right)
^{2n+1-2j}}{j!\left( 2n+1-2j\right) !}%
\end{array}%
\begin{array}{c}
\mathrm{for}\text{ }\psi \text{ \textrm{even}} \\
\\
\mathrm{for}\text{ }\psi \text{ \textrm{odd.}}%
\end{array}%
\right.
\end{equation}%
Then, using the formula for the Hermite polynomial \cite{leb}%
\begin{equation}
H_{n}\left( y\right) =n!\sum_{j=0}^{\left[ \frac{n}{2}\right] }\frac{\left(
-1\right) ^{j}\left( 2y\right) ^{n-2j}}{j!\left( n-2j\right) !},  \label{her}
\end{equation}%
where $\left[ n/2\right] $ denotes the largest integer $\leq n/2$, the
eigenfunction is found to be%
\begin{equation}
\psi _{n}\left( x\right) =\frac{c_{0}^{\left( n\right) }}{\sqrt{\pi }}\exp
\left( -\frac{m\omega }{2\hslash }x^{2}\right) \times \left\{
\begin{array}{c}
\frac{\Gamma \left( n+1/2\right) }{\left( 2n\right) !}\,H_{2n}\left( \sqrt{%
\frac{m\omega }{\hslash }}x\right) \\
\\
\frac{n!}{\left( 2n+1\right) !}\,H_{2n+1}\left( \sqrt{\frac{m\omega }{%
\hslash }}x\right)%
\end{array}%
\begin{array}{c}
\mathrm{for}\text{ }\psi \text{ \textrm{even}} \\
\\
\mathrm{for}\text{ }\psi \text{ \textrm{odd.}}%
\end{array}%
\right.
\end{equation}%
Because $H_{n}\left( -y\right) =\left( -1\right) ^{n}H_{n}\left( y\right) $,
the solution can also be expressed as%
\begin{eqnarray}
E_{n} &=&\left( n+\frac{1}{2}\right) \hslash \omega ,\quad n=0,1,2,\ldots
\notag \\
&& \\
\psi _{n}\left( x\right) &=&A_{n}\,\,e^{-m\omega x^{2}/\left( 2\hslash
\right) }\,H_{n}\left( \sqrt{\frac{m\omega }{\hslash }}x\right) ,  \notag
\end{eqnarray}%
where $A_{n}$ are normalization constants.

\section{Conclusion}

We have shown that the complete solution of the one-dimensional quantum
harmonic oscillator can be approached via the Laplace transform method with
simplicity and elegance, even if the eigenfunction does not vanish at the
origin. Englefield%
\'{}%
s recipe concedes to explore asymptotic expansions of the eigenfunction and
its Laplace transform to gain information from the Schr\"{o}dinger equation
even if it is not possible to solve it in closed form. The discussion
presented here may apply to any other problem that, after factorizing the
behaviour at the origin and at infinity, reduces to the confluent
hypergeometric equation such as the pseudoharmonic, Coulomb and Kratzer
potentials.

\begin{acknowledgments}
This work was supported in part by means of funds provided by CNPq and FAPESP.
\end{acknowledgments}


\begin{thebibliography}{99}
\bibitem{f1} The series expansion about $s=1$ would ruin the good behaviour
of $\phi \left( \xi \right) $ as $\xi \rightarrow \infty $.
\bibitem{f2} The quantization condition is the root of the indicial
condition.
\end{thebibliography}
\end{document}